\documentclass[twocolumn]{aastex62}

\usepackage{graphics,epsf}
\usepackage{amsmath}                
\usepackage{amsfonts}               
\usepackage{amssymb}                
\usepackage{epsfig}                 
\usepackage{graphicx}
\usepackage{float}
\usepackage{color}
\usepackage[para,online,flushleft]{threeparttable}

\def \cm{~\rm{cm}}
\def \s{~\rm{s}}
\def \km{~\rm{km}}

\def \erg{~\rm{erg}}

\begin{document}

\title{Possible indications for jittering jets in core collapse supernova explosion simulations}



\author[0000-0003-0375-8987]{Noam Soker}
\affiliation{Department of Physics, Technion, Haifa, 3200003, Israel; soker@physics.technion.ac.il}
\affiliation{Guangdong Technion Israel Institute of Technology, Shantou 515069, Guangdong Province, China}


\begin{abstract}
I analyse recent three-dimensional hydrodynamical simulations of exploding core collapse supernovae (CCSNe) and find that a rapid increase in the diagnostic explosion energy occurs parallel to, or very shortly after, the accretion of gas with relatively large amount of angular momentum and/or specific angular momentum. The tendency towards a positive correlation between angular momentum accretion and the increase in the explosion energy appears despite the absence of magnetic fields in the simulations. I speculate that the inclusion of magnetic fields would make this positive correlation much stronger and would increase the explosion energy. At times, the high angular momentum accretion leads to a bipolar outflow that in some cases changes directions. 
I argue that there is a mutual influence between neutrino heating and stochastic angular momentum, and that both operate together to explode the star with bipolar outflows that change directions, namely, jittering jets. 
This study calls for a deep flow analysis of simulations of CCSNe, e.g., to present the flow structure, the density, and the angular momentum of the flow near the newly born neutron star (NS).  
\end{abstract}



\section{Introduction}
\label{sec:intro}

There are two basic theoretical mechanisms to utilise the gravitational energy of the collapsing core of core collapse supernovae (CCSNe) to power the explosion. These are the delayed neutrino mechanism \citep{BetheWilson1985} and the, 25 years younger, jittering jets explosion mechanism (\citealt{Soker2010}; or more generally the jet feedback mechanism, e.g., \citealt{Soker2016Rev}).  
A pure delayed neutrino mechanism, where the heating by neutrino revives the stalled shock to obtain the desired explosion energy, seems to have problems (e.g., \citealt{Papishetal2015, Kushnir2015b}). These problems require the addition of some ingredients to the delayed neutrino mechanism. As well, numerical simulations find no stochastic accretion disks around the newly born neutron stars (NSs), something that requires adding ingredients to the jittering jets explosion mechanism. 
 
The extra ingredient that recent studies suggest to facilitate the jittering jets explosion mechanism is heating by neutrinos 
\citep{Soker2018KeyRoleB, Soker2019SASI}. 
This comes on top of the main ingredients of the jittering jets explosion mechanism that are (e.g., \citealt{Soker2017RAA}) the flow fluctuations in the convective regions of the pre-collapse core or envelope \citep{GilkisSoker2014, GilkisSoker2015, Quataertetal2019} that reach the stalled shock at about $150 \km$ with relatively large amplitudes (e.g., \citealt{AbdikamalovFoglizzo2019}), and the spiral standing accretion shock instability (SASI) and other instabilities that amplify the flow fluctuations behind the stalled shock and above the newly born NS or black hole as the mass flows in (for the spiral SASI itself see, e.g., \citealt{BlondinMezzacappa2007, Iwakamietal2014, Kurodaetal2014, Fernandez2015, Kazeronietal2017}). These two ingredients lead to accretion flow with stochastic varying angular momentum, both in magnitude and in direction. These in turn lead, according to the jittering jets explosion mechanism, to the launching of jittering jets, but only with the extra ingredients of magnetic fields and/or extra heating by neutrinos (e.g., \citealt{Soker2019Dynamo} that lists the  unique properties of the jittering jets explosion mechanism). 

The main extra ingredient that recent studies concentrate on to overcome the difficulties of the delayed neutrino mechanism is the introduction of flow fluctuations in the pre-collapse core (e.g., \citealt{CouchOtt2013, CouchOtt2015, MuellerJanka2015, Mulleretal2017}). These flow fluctuations occur in the convective regions above the iron inner core. 
These fluctuations are the main ingredient of the jittering jets explosion mechanism, and, as mentioned above, lead to the accretion of gas with stochastic angular momentum onto the newly born NS or black hole (e.g., \citealt{GilkisSoker2015}). 
  
Some recent three-dimensional (3D) numerical simulations of CCSNe show that the diagnostic explosion energy increases over a long time as the central newly born NS continues to blow non-spherical winds (section \ref{sec:jittering}). In this study I examine (section \ref{sec:jittering}) whether these long-lived outflows have some properties that might hint on the development of jittering jets.
Although these simulations do not include magnetic fields that researchers usually assume to be necessary for launching jets (e.g., review by \citealt{Livio1999}), there are theoretical mechanisms to launch jets that do not require magnetic fields. These mechanisms rather use thermal pressure (e.g., \citealt{TorbettGilden1992}), e.g., as a result of shocks in accretion disks around young stars (e.g., \citealt{SokerRegev2003}) and white dwarfs (e.g., \citealt{SokerLasota2004}). As well, there are mechanisms for jet launching that do not require a fully developed accretion disk but do require magnetic fields (e.g., \citealt{SchreierSoker2016}). 
I summarise my tentative findings in section \ref{sec:summary}. 

\section{Analysing numerical simulations}
\label{sec:jittering}

\subsection{Pre-collapse perturbations}
\label{subsec:Muller2017}

I analyse the results of \cite{Mulleretal2017} who introduce large scale perturbations in the pre-collapse core of a model with zero age main sequence mass of $M_{\rm ZAMS}=18 M_\odot$. 
These perturbations were introduced only in the mass zone $1.68 M_\odot < m <  4.07 M_\odot$.  
In an earlier paper (\citealt{Soker2018KeyRoleB} where I present analysis of other aspects of the simulations of \citealt{Mulleretal2017} that I will not repeat here) I commented that I expect perturbations at smaller mass coordinates (e.g., \citealt{Zilbermanetal2018}). This implies that accretion of gas with stochastic angular momentum might start earlier than what \cite{Mulleretal2017} find, with more pronounced effects on the outflow. 

\begin{figure}
\hskip -1.50 cm
\includegraphics[width=0.62\textwidth]{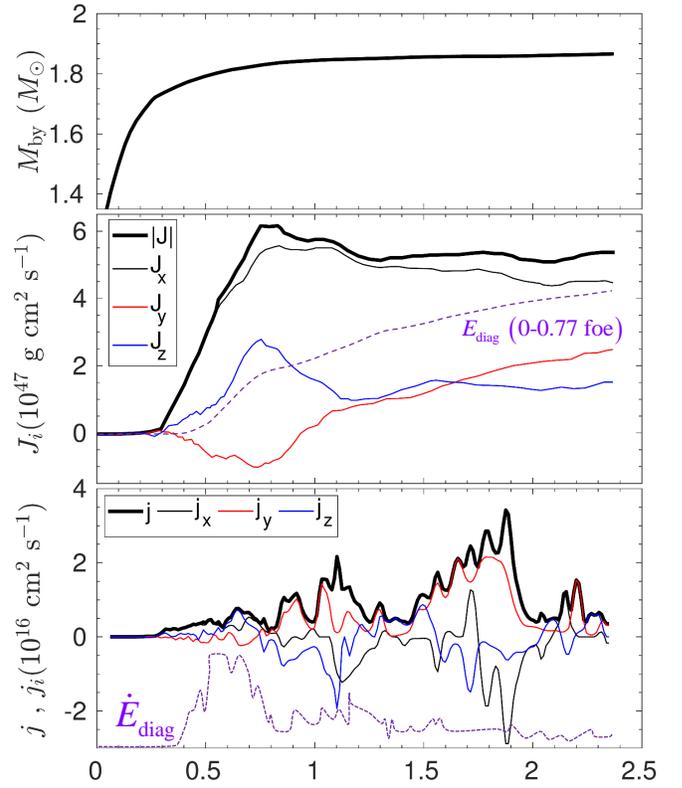}
\vskip -2.00 cm
\caption{The angular momentum of the accreted mass in a simulation performed by \cite{Mulleretal2017} that includes pre-collapse perturbations. The two upper panels are from figure 20 there. The upper panel shows the baryonic mass of the neutron star and the second panel shows the components of the angular momentum of the neutron star and its total angular momentum. The dashed purple line in the second panel is the diagnostic explosion energy $E_{\rm diag}$ from their figure 15 (its lower value is $0$ and its maximum value $0.77~$foe, where foe$\equiv10^{51}\erg$). The lower panel is the specific angular momentum of the accreted gas that I calculated from the two upper panels by $j_i=dJ_i/dM_{\rm by}$ and $j$ is the total specific angular momentum. The dashed purple line is the rate of change of the diagnostic energy, $\dot E_{\rm diag}$, that I smoothed from their figure 17 (here in relative units with a minimum value of zero). 
Note the high value of $\dot E_{\rm diag}$ in the time period $t \simeq 0.4-0.8 \s$ that occurs at the same time of the large increase in the angular momentum of the NS (black line in second panel). Another possible positive correlation is between the high specific angular momentum of the accreted mass and a large increase in diagnostic energy in the time period $t\simeq 0.9-1.3 \s$ }
\label{fig:Muller2017}
\end{figure}

I present some quantities of their run with large pre-collapse perturbations in Fig. \ref{fig:Muller2017}. 
In the upper two panels I present the baryonic mass $M_{\rm by}$ of the NS and the angular momentum of the NS, total angular momentum $J$ and components $J_x$, $J_y$ and $J_z$, from their figure 20.
In the second panel I added the evolution of the diagnostic explosion energy $E_{\rm diag}$ from their figure 15.  
From these two panels I calculated the specific angular momentum of the accreted gas by  $j_i=dJ_i/dM_{\rm by}$, where $i=x,y,z$, and then $j=\left( j^2_x+j^2_y+j^2_z \right)^{1/2}$. 
I plot these quantities in the lower panel, as well as the rate of change of the diagnostic energy that I take from their figure 17. 

\cite{Mulleretal2017} argue that the rate of change of the diagnostic energy $\dot E_{\rm diag}$ correlates with the neutrino luminosity and with the nucleon recombination power. Their figure 17 shows some correlation, but not a perfect one. Here I point out that there is also some correlation between large changes in the angular momentum accretion rate (by components) and the magnitude of $\dot E_{\rm diag}$, as evident from Fig. \ref{fig:Muller2017}. In particular correlation exists in the time period $0.4 \s \la t \la 0.8 \s$ when there is a large increase in the angular momentum of the NS, and in the time period $0.9 \s \la t \la 1.3 \s$ when the magnitude of the specific angular momentum is large. In both these periods $\dot E_{\rm diag}$ is relatively large.

The typical value of the amplitude of the specific angular momentum fluctuations is $j_A \simeq 10^{16} \cm^2 \s^{-1}$. This typical value is a large fraction of the angular momentum for a Keplerian orbit around the NS 
\begin{eqnarray}
j_{\rm Kep} 
=\left( G M_{\rm NS}r \right)^{1/2} &
\nonumber  \\ =  
2 \times 10^{16} 
\left( \frac{M_{\rm NS}}{1.5 M_\odot} \right)^{1/2}  &
\left( \frac{r}{20 \km} \right)^{1/2} \cm^2 \s^{-1}. 
\label{eq:jKep} 
\end{eqnarray}
This suggests that the angular momentum might play an important role in determining the outflow geometry from the NS vicinity, i.e., forming a bipolar outflow (more in section \ref{subsec:bipolar}). 

In principle it is possible that the correlation between angular momentum accretion rate and the diagnostic explosion energy results from the explosion process that determines the accretion process. However,  \cite{Mulleretal2019Jittering} note that
the NS reaches its almost final spin before the onset of the explosion. It seems therefore that angular momentum fluctuations do play a significant role in setting the explosion and its general bipolar geometry. 
I take the view that there is a mutual influence between angular momentum and neutrino heating, as I explain in more detail in section \ref{subsec:bipolar}. 

In the time period $1.55 \s \la t \la 1.85 \s$, i.e., for about $0.3 \s$, the specific angular momentum component $j_y$ has a large value of $j_y \ga 10^{16} \cm^2 \s^{-1}$. Inclusion of magnetic fields might lead to the launching of two opposite jets even if a fully Keplerian accretion disk does not form \citep{SchreierSoker2016}. The NS accretes a mass of $\approx 0.005 M_\odot$ during that time period. If the high-angular momentum flow launches jets that carry a fraction of $0.2$ of the accreted mass at the escape speed from the newly born NS, then the energy in the jets is $\approx 10^{50} \erg$. This is about 10 percent of the explosion energy. Two opposite jets that maintain their axis and carry  $\approx 1-30 \%$ of the explosion energy might explain the formation of two opposite protrusions, termed `ears', in some supernova remnants (\citealt{Bearetal2017, GrichenerSoker2017, YuFang2018}). 
The results of the 3D simulations of \cite{Mulleretal2017} hint into the possibility that the NS might launch jets at the end of the explosion process. 
  
The kinetic energy due to the angular momentum of the accreted gas might crudely indicate the possible importance of including magnetic fields. Turbulence and shear due to angular momentum, i.e., shear due to rotational velocity that is perpendicular to both the angular momentum axis and to the radial direction, as well as the stochastic nature of the angular momentum axis \citep{Soker2019Dynamo}, can amplify magnetic fields near the NS. I here consider only the rotational velocity due to angular momentum itself, hence this treatment underestimates the energy available from the accretion flow. 
 
I take the accreted mass to be that from $t=0.5 \s$ for the simulation presented in Fig. \ref{fig:Muller2017}. This mass  is $M_{\rm acc,f}\simeq 0.05M_\odot$. The lower panel of Fig. \ref{fig:Muller2017} shows that the specific angular momentum has a typical amplitude of $A_j \simeq 10^{16} \cm^2 \s^{-1}$. For an accretion onto a newly born NS with a radius 
of $R_{\rm NS}=20 \km$, the kinetic energy due to the rotational velocity is 
\begin{eqnarray}
E_{\rm rotation} \simeq  
\frac{1}{2} M_{\rm acc,f} \left(\frac{A_j}{R_{\rm NS}} \right)^2 \approx  10^{51}  &
\left( \frac{M_{\rm acc,f}}{0.05 M_\odot} \right) 
\nonumber  \\  \times
  \left( \frac{A_j}{10^{16} \cm^2 \s^{-1}} \right)^2  
 \left( \frac{R_{\rm NS}}{20 \km} \right)^{-2}  &
 \erg. 
\label{eq:Erot} 
\end{eqnarray}
Even magnetic activity cannot channel all this energy to an outflow. On the other hand, there is a shear due to the radial inflow of the accretion gas that can further amplify the magnetic fields \citep{Soker2018KeyRoleB}, and, as mentioned above, there is the turbulence and stochastic angular momentum. The main point to take from equation (\ref{eq:Erot}) is my expectation  that the inclusion of magnetic fields in future numerical simulations will lead to more energetic explosions and to clearer formation of jets. 

\subsection{The appearance of a bipolar outflow}
\label{subsec:bipolar}

\begin{figure}
\hskip -2.00 cm
\includegraphics[trim= 0.0cm 0.0cm 0.0cm 0.0cm,clip=true,width=0.65\textwidth]{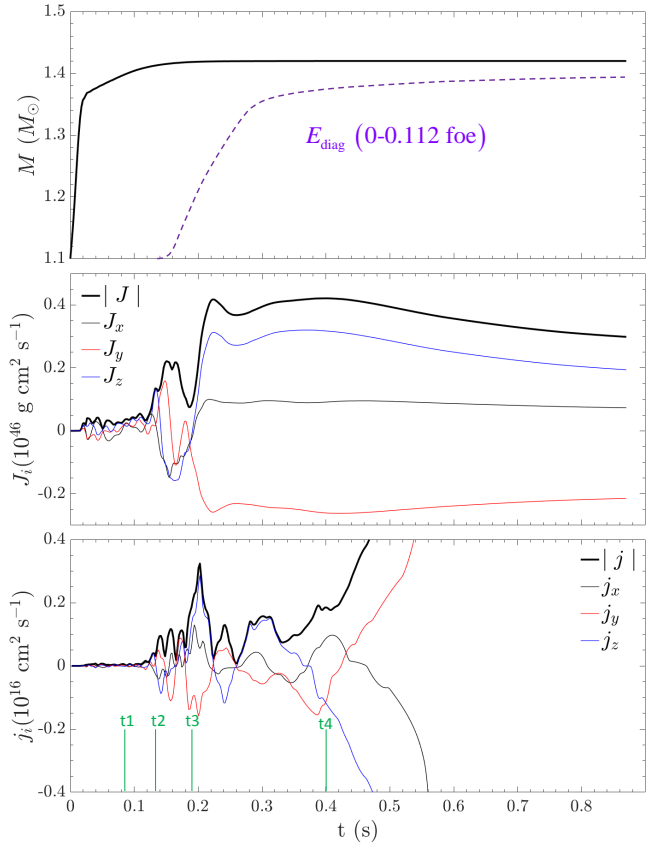}
\vskip -2.500 cm
\caption{Results from a simulation performed by \cite{Mulleretal2018Bipol} of the collapse of a striped star (no hydrogen and only little helium at collpase). This model has no pre-collaפse perturbations, therefore the values of the stochastic accreted angular momentum are smaller than in Fig. \ref{fig:Muller2017}.
The two upper panels are from figure 5 there. The upper panel shows the baryonic mass of the NS with the addition of the diagnostic explosion energy in dashed-purple line from their figure 2 (its lower value is $0$ and its maximum value is $0.112~$foe, where foe$\equiv10^{51}\erg$). The second panel shows the components of the angular momentum of the neutron star and its total angular momentum. 
The lower panel is the specific angular momentum of the accreted gas that I calculated from the two upper panels (see Fig. \ref{fig:Muller2017}). 
The green vertical lines mark the times of the four panels in Fig. \ref{fig:Muller2018B}. 
This figure emphasises the following. (1) The bipolar outflow that is seen in the lower left panel of Fig. \ref{fig:Muller2018B}, marked here as $t=t3$, occurs during the time of rapid increase in the angular momentum of the NS. Namely, during the accretion phase of high angular momentum gas. (2) The large increase in diagnostic explosion energy comes alongside with the high-angular momentum accretion phase.   } 
\label{fig:Muller2018}
\end{figure}

\begin{figure}
\hskip -2.50 cm
\includegraphics[trim= 0.0cm 4.0cm 0.0cm 0.0cm,clip=true,width=0.75\textwidth]{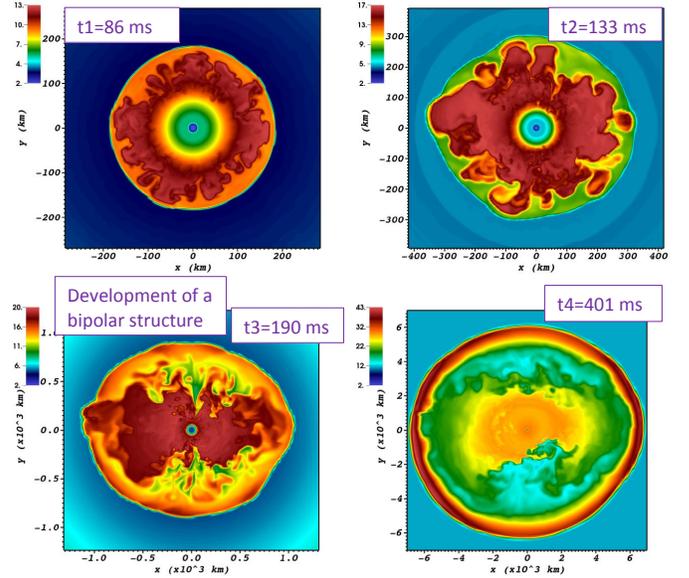}
\vskip -5.00 cm
\caption{Results from a simulation performed by \cite{Mulleretal2018Bipol} of the collapse of a striped star. The panels show the entropy maps in the $xy$ plane at four times (also indicated by vertical green lines in the lower panel of Fig. \ref{fig:Muller2018}). Units on the axes are in $1000 \km$ and entropy in units of $k_{\rm b}$/nucleon.}
\label{fig:Muller2018B}
\end{figure}

In this section I examine the results of \cite{Mulleretal2018Bipol} who study the explosion of a striped stellar model, i.e., no hydrogen and only a small mass of helium in the envelope of the star at core collapse.   In Fig. \ref{fig:Muller2018} I present the results of their 3D simulation in the same manner as I did in Fig. \ref{fig:Muller2017}. In Fig. \ref{fig:Muller2018B} I present their figure 3 that shows the entropy maps in a plane at four times (entropy in units of $k_{\rm b}$/nucleon). 
 
The high entropy regions indicate outflows. 
The clear bipolar structure of the regions with high-entropy that appears in the third panel of Fig. \ref{fig:Muller2018B} comes shortly after the specific angular momentum of the accreted gas increases (time t3 in the lower panel of Fig. \ref{fig:Muller2018}).
The axis of the bipolar structure on the $xy$ plane is along the $y$-axis. This is expected in the frame of the jittering jets explosion mechanism because of $j_x<j_y$. 

As in the case of the flow that I study in section \ref{subsec:Muller2017}, there is one of three possibilities. (1) The neutrino heating alone determines the outflow geometry that in turn determines the angular momentum of the accreted gas. (2) Only the angular momentum of the accreted gas determines the outflow geometry. (3) There is an interplay between angular momentum of the accreted gas and neutrino heating. 
 
I suggest the third possibility. Even in the case of pure jet-core interaction  one episode of launching jets influences the angular momentum of the accreted gas of later episodes by preventing accretion along the direction of the jets (e.g., \citealt{PapishSoker2014Planar}). So instabilities that lead to directional mass outflow by neutrino heating can influence the angular momentum of later accreted gas. 

At the same time angular momentum helps mass ejection along the angular momentum axis even if a Keplerian disk does not form. This is because the inflow along the angular momentum axis is less dense. A parcel of gas with angular momentum $j$ around an axis $\vec r$ cannot be accreted within an angle $\theta_a$ from the axis $\vec r$, given by \citep{Papishetal2016, Soker2018KeyRoleB}
\begin{eqnarray}
\theta_a = &
\sin^{-1} \sqrt{\frac{j(t)}{j_\mathrm{Kep}}} \simeq 
10^\circ \left( \frac{M_\mathrm{NS}}{1.5 M_\odot} \right)^{-1/4}
\\ & \times
\left( \frac{r}{50 \km} \right)^{-1/4} 
\left( \frac{j(t)}{10^{15} \cm^2 \s^{-1}} \right)^{1/2},
 \nonumber
\label{eq:angle}
\end{eqnarray}
where the second equality holds for small angles.  

It is crucial to understand that the important role of neutrino heating does not preclude the role of jittering jets. The general heating by a central energy source, which here is neutrino heating, is not new in some mechanisms for launching jets. \cite{Livio1999} mentions in his review that ``the production of powerful jets requires a hot corona or access to an additional energy source associated with the central object.'' 
The most extreme type of central energy source might be spinning supermassive black holes that accrete mass and launch jets with energies larger than the gravitational energy that the accreted mass releases. \cite{McNamaraetal2009}, for example, suggested that the central supermassive black hole of the cluster of galaxies  MS0735.6+7421 released about $10^{62} \erg$ of its spinning energy in launching jets. These jets inflated the huge bipolar bubbles in the hot intracluster gas of this cluster. Channelling of the spin energy of a black hole to a bipolar outflow (two opposite jets) requires the presence of magnetic fields (e.g., \citealt{Nemmenetal2007}). 
It is possible that in CCSNe the neutrino heating is the additional energy source for forming jets, that acts on top of the non zero angular momentum of the accreted gas. To utilise the neutrino energy there is no need for magnetic fields.  
  
I note that the specific angular momentum fluctuations in the simulation that Fig. \ref{fig:Muller2018} and \ref{fig:Muller2018B} present are small, but so is the explosion energy of $E_{\rm diag} \simeq 10^{50} \erg$. Larger explosion energies, as in the simulation that Fig. \ref{fig:Muller2017} presents, require larger angular momentum fluctuations. From Fig. \ref{fig:Muller2017} I find  that the avoided angle for that simulation reaches values of $\theta_a \simeq 30^\circ$.  
     
\subsection{Signatures of jittering jets}
\label{subsec:jitteringjets}
  
\begin{figure}
\hskip -4.00 cm
\includegraphics[trim= 0.0cm 0.0cm 0.0cm 0.0cm,clip=true,width=0.75\textwidth]{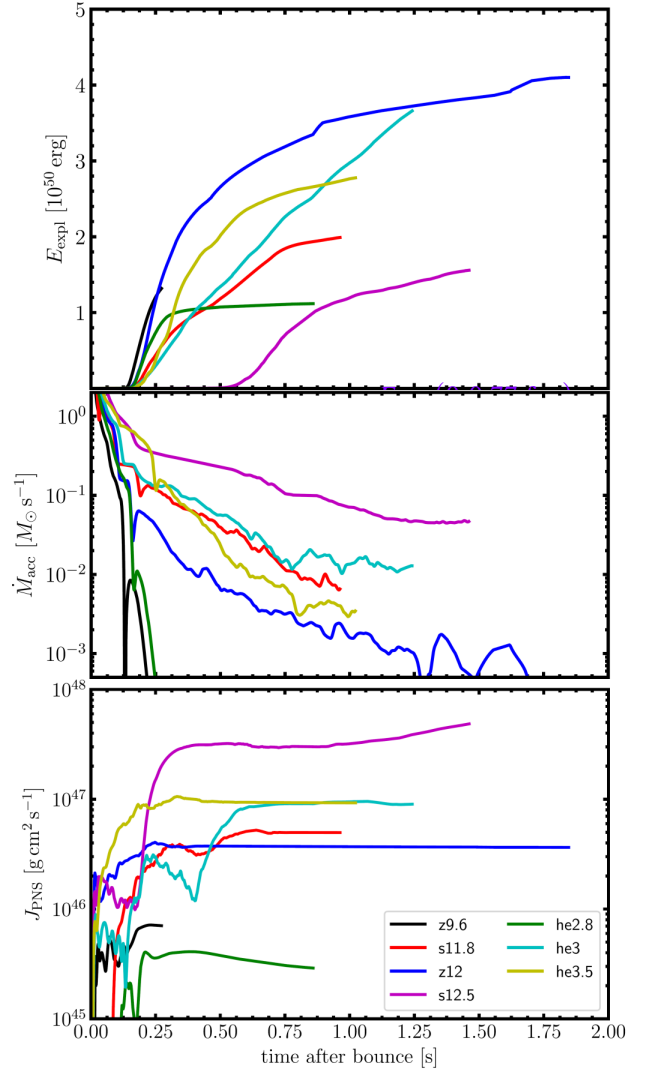}
\vskip -1.00 cm
\caption{Results of simulations of seven models from  \cite{Mulleretal2019Jittering}. The upper, middle, and lower panels show the diagnostic explosion energy, the mass accretion rate, and the angular momentum of the NS, respectively. 
Note the appearance of angular momentum fluctuations before the explosion energy becomes positive, and the large and rapid variations of the angular momentum that occur near the time when the explosion energy becomes positive and then rapidly increases.  }
\label{fig:Muller2019A}
\end{figure}

I examine some of the results of \cite{Mulleretal2019Jittering} who conducted 3D simulations of seven models, four that evolve as single stars and suffer little mass loss (z9.6, s11.8, z12, and s12.5), and 3 models where the progenitor transfers mass to a companion and evolves to become a stripped CCSN (he2.8, he3.0, he3.5). In four models (s11.8, z12, s12.5, and he3.0) they introduce pre-collapse perturbations above the iron core due to oxygen burning. I present some properties of the seven model in Fig. \ref{fig:Muller2019A}.

Only model s12.5 acquires positive $E_{\rm diag}$ at a late time. 
\cite{Mulleretal2019Jittering} attribute this to the explosion geometry. 
They notice that all six models beside s12.5 exhibit, in their words,  ``some degree of bipolarity at the early stages of the explosion, either with two similarly prominent outflows (model z12) or with a strong and a subdominant outflow in the opposite.'' 
In model s12.5, they notice, the explosion is clearly unipolar from early times.
In other astrophysical objects, from young stellar objects through planetary nebulae and active galactic nuclei (e.g., \citealt{Livio1999}), such bipolar outflows are termed `jets'. It is important to note that such outflows can be wide,  e.g., in active galactic nuclei (e.g., \citealt{Dunnetal2012}), but they still have the physics of jets, and hence I term them jets (for a review see \citealt{Soker2016Rev}).  

\begin{figure}
\hskip -2.00 cm
\includegraphics[trim= 0.0cm 5.0cm 0.0cm 0.0cm,clip=true,width=0.65\textwidth]{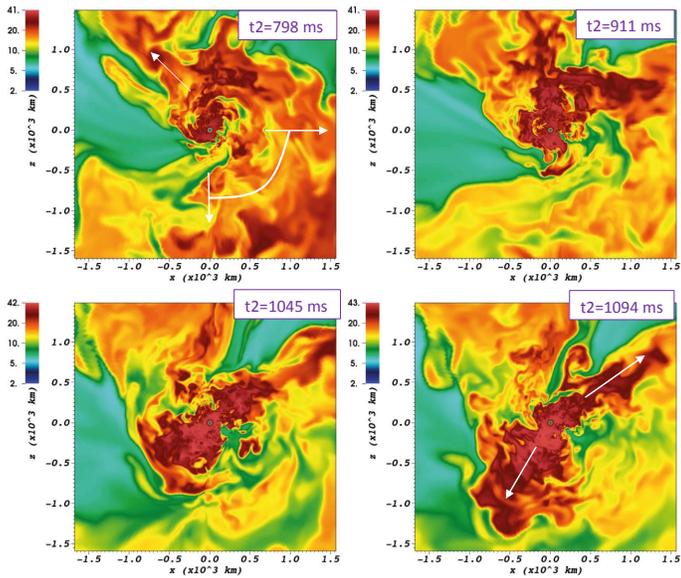}
\vskip -3.00 cm
\caption{The entropy maps in the $zx$ plane from simulation s11.8 of \cite{Mulleretal2019Jittering} at 4 times. Units on the axes are in $1000 \km$. In this model they introduced pre-collapse perturbations in the core. I added the white arrows to indicate the high entropy regions that indicate outflows. Note the structure of a bipolar outflow that changes its direction from the first panel to the last. This is the expected behavior in the jittering jets explosion mechanism. 
 }
\label{fig:Muller2019B}
\end{figure}

\begin{figure}
\includegraphics[trim= 0.0cm 0.0cm 0.0cm 0.0cm,clip=true,width=0.45\textwidth]{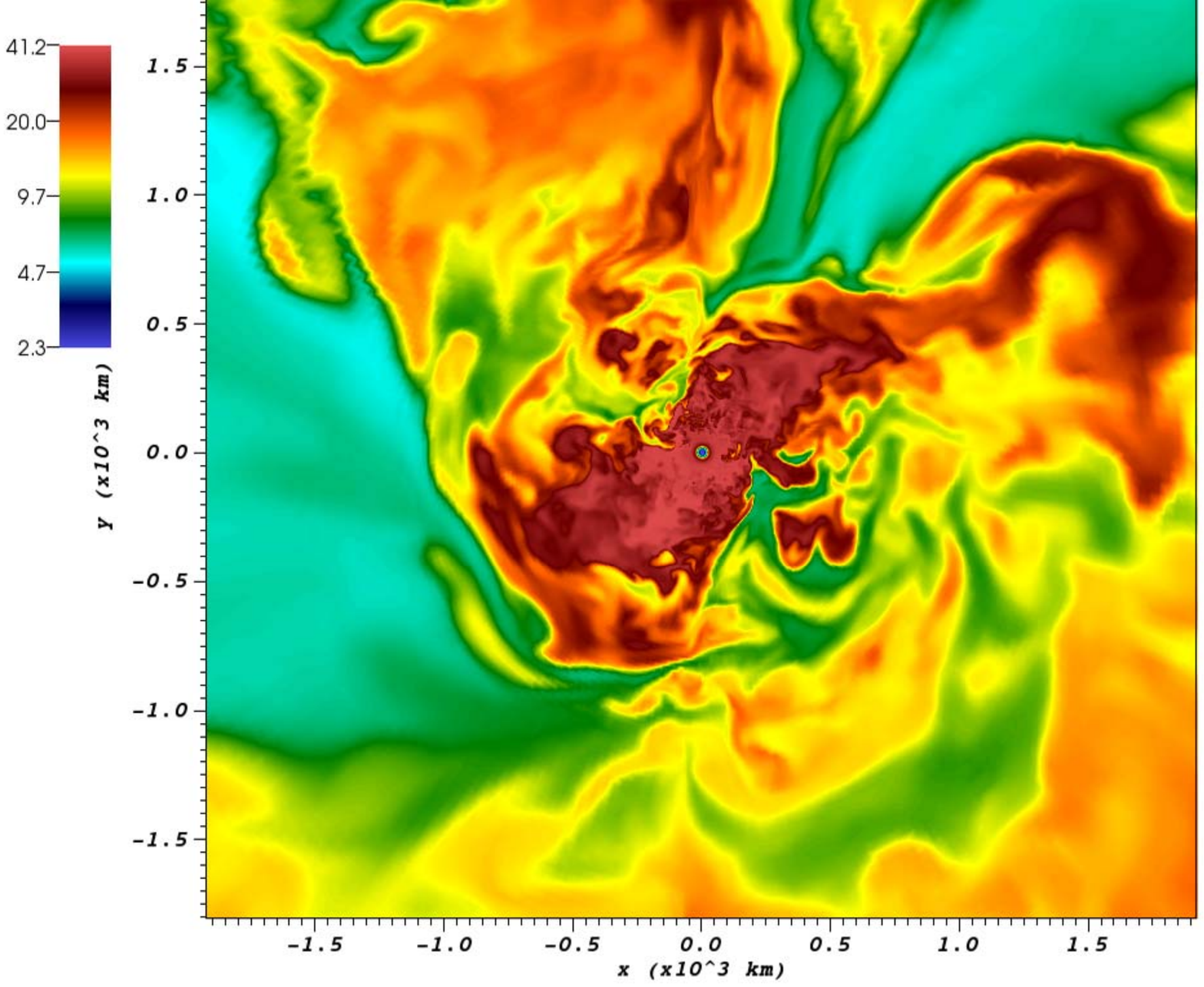}
\includegraphics[trim= 0.0cm 0.0cm 0.0cm 0.0cm,clip=true,width=0.45\textwidth]{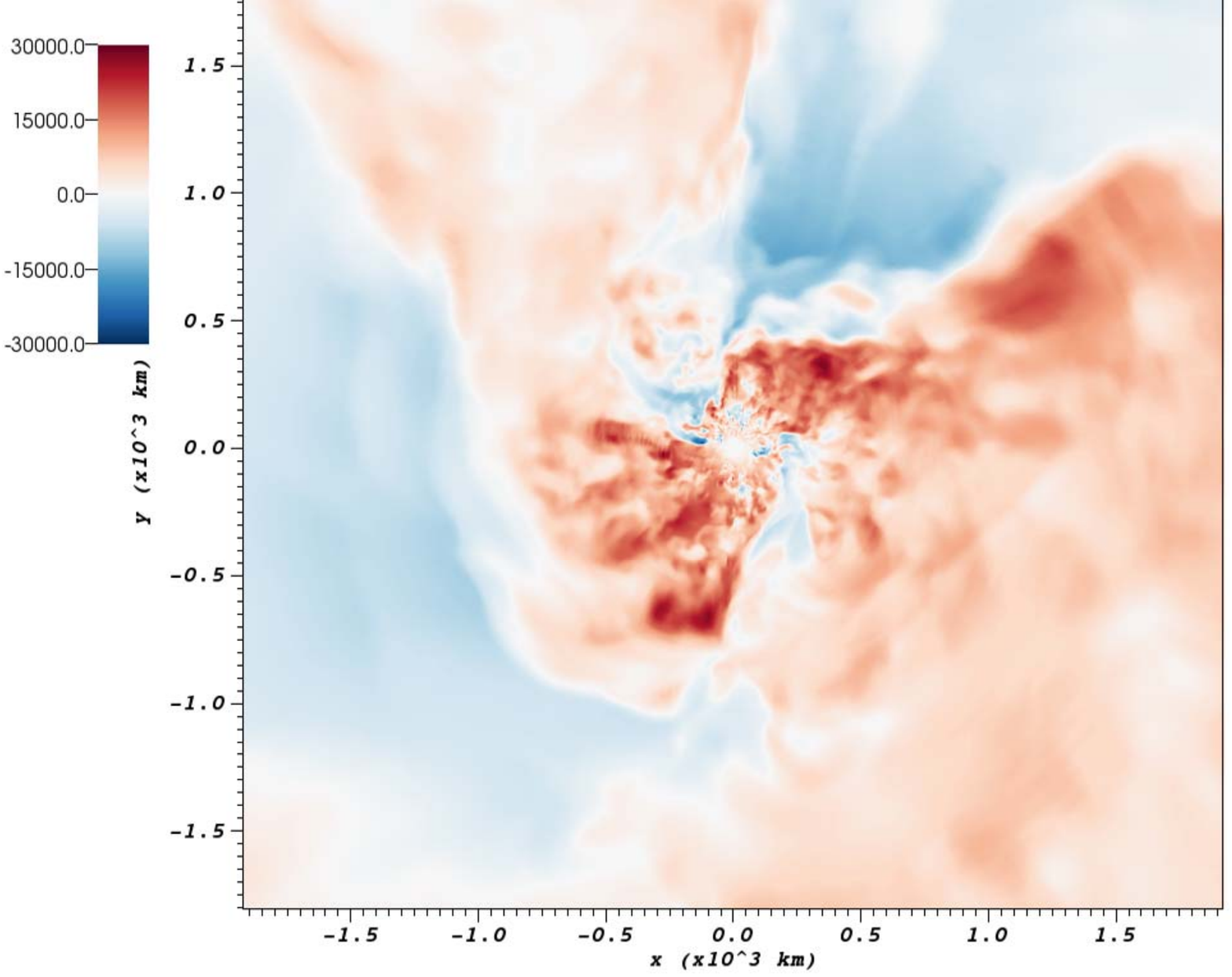}
\caption{The entropy map (upper panel) and radial velocity map (lower panel; from $v_r=-3\times 10^4 \km \s^{-1}$ to $v_r=-3\times 10^4 \km \s^{-1}$), for the same run and same plane as in Fig. \ref{fig:Muller2019B} but at $t=1037~$ms (kindly supplied to the author by Bernhard  M{\"u}ller). (Note that despite the labelling the plane is as in Fig. \ref{fig:Muller2019B}.)
This figures shows that the high entropy regions indicate also an outflow.  
 }
\label{fig:Muller2019C}
\end{figure}

The results of \cite{Mulleretal2019Jittering} reveal more than the importance of the  bipolar outflow. They reveal, what I interpret as, jittering jets. In Fig. \ref{fig:Muller2019B} I present the entropy maps in the $zx$ plane for their model s11.8 at four times, as they present in their figure 7. The high entropy regions indicate outflows, as the radial velocity map and entropy map in Fig. \ref{fig:Muller2019C} show. The outflow in Fig. \ref{fig:Muller2019B} has a bipolar structure in the first panel, and a bipolar structure in the last panel, but along different directions as I marked by the white arrows. This change in the direction of a bipolar outflow over a time scale of $\approx 0.1 \s$ is the patter that signifies the jittering jets explosion mechanism
\citep{PapishSoker2011}. 

I take it therefore that 3D simulations of CCSNe do obtain jittering jets, and that neutrino heating does play a significant role at low explosion energies of $E_{\rm diag} \la 10^{51} \erg$.  My expectation is that the significance of neutrino heating decreases with increasing explosion energies beyond $\simeq 10^{51} \erg$, when magnetic effects become more and more important. 
 
\cite{Mulleretal2019Jittering} find that there is a suggestive trend for the angle between the NS final spin and the direction of the natal kick velocity to cluster at $\alpha > 50^\circ$. This is a very interesting result in light of the finding by \cite{BearSoker2018kick}. \cite{BearSoker2018kick} infer the direction of the two opposite jets that shape, according to the jittering jets explosion mechanism, the axisymmetrical morphological features that appears in 12 SN remnants that also have measured NS kick velocity. They then correlate the jets'-axis direction with the NS natal kick direction. They find that the distribution is almost random, but missing small angles. The finding of \cite{Mulleretal2019Jittering} is compatible with the finding of \cite{BearSoker2018kick}. This suggests that the newly born NS does launch jets. I speculate that the inclusion of magnetic fields in numerical simulations will eventually lead to the launching of jets that can leave axi-symmetrical imprints in some SN remnants, as observed. 

\section{Discussion and Summary}
\label{sec:summary}

I analysed three papers \citep{Mulleretal2017, Mulleretal2018Bipol, Mulleretal2019Jittering} that present 3D hydrodynamical simulations of CCSNe. 
I found that in all cases accretion of mass with stochastic variations of angular momentum, magnitude and direction, has some correlation with the onset of explosion and the increase of explosion energy. 

On average, this correlation, although not perfect, is clearer in simulations that have pre-collapse perturbations due to convective burning zones. 
For example, in Fig. \ref{fig:Muller2017} (from \citealt{Mulleretal2017}) the increase rate of the explosion energy, $\dot E_{\rm diag}$, has large values when the angular momentum of the NS increases and then when the specific angular momentum of the accreted mass is large (lower panel of Fig. \ref{fig:Muller2017}).
 
Fig. \ref{fig:Muller2018} (from \citealt{Mulleretal2018Bipol}) shows that even when there are no pre-collapse perturbations the increase in explosion energy occurs when the accreted mass has a relatively large specific angular momentum.
In that simulation a clear bipolar outflow occurs (third panel of Fig. \ref{fig:Muller2018B}) around the time when the specific angular momentum of the accreted gas is relatively high. 

Fig. \ref{fig:Muller2019B} (from \citealt{Mulleretal2019Jittering}) shows the entropy maps of one simulation from the seven simulations that Fig. \ref{fig:Muller2019A} presents. Outflows take place in the high entropy regions (Fig. \ref{fig:Muller2019C}). As I marked with white arrows, the bipolar axis changes its direction from the first to last panel. I take it to imply the presence of jittering jets.   

My conclusion is that there is an interplay between angular momentum of the accreted gas and neutrino heating, and that both act together to explode the star with varying bipolar outflows, i.e., jittering jets.  This is the case even when the net angular momentum of the pre-collapse core is zero. It is important to emphasise that mechanisms to launch jets without magnetic fields are not new (e.g., \citealt{TorbettGilden1992}), and so is the old idea that in some cases the central object supplies extra energy to facilitate the launching of jets (e.g., review by \citealt{Livio1999}). In the present setting neutrinos carry the extra energy. I also note that jets can be wide, i.e., with a half opening angle of tens of degrees, but still maintain the properties of more collimated jets, e.g., a bipolar morphology, even if the two sides are not equal, hence I term them jets.   
 
To better reveal these features of angular momentum variations and jittering bipolar outflows (jets), analysis of simulations of CCSNe should perform a deep flow analysis, e.g., to present the flow structure, the density, and the angular momentum of the flow near the newly born NS (or black hole). 
 
I raise few more points that might follow from the conclusion that there is a mutual influence between neutrino heating and angular momentum and that both operate together to explode the star with jittering jets.     

(1) The present conclusion implies that 1D models cannot reproduce the full explosion mechanism, even if they incorporate some effects of convection (e.g., \citealt{Couchetal2019, Mabantaetal2019}). 

(2) The Crab Nebula has a pulsar with a well define rotation axis. Estimates put the kinetic energy of the explosion at $\approx 10^{50} \erg$ (e.g., \citealt{YangChevalier2015}). This low explosion energy might require no jets. However, the remnant itself has a  protrusion, an `ear', along the rotation axis of the pulsar with a morphology that suggests shaping by a jet that carried an energy of $\approx 3 \%$ of the total CCSN kinetic energy \citep{GrichenerSoker2017}. This jet might be from the last jet-launching episode of several \citep{Bearetal2017}.  The same holds for several other supernova remnants with pulsars that show ears that hint at shaping by jets   \citep{Bearetal2017, GrichenerSoker2017}. In the Vela SN remnant the symmetry axis of the Si-rich jet-counterjet structure that \cite{Garciaetal2017} infer has a different direction than the line connecting the two ears \citep{GrichenerSoker2017}, hinting on jittering jets. 

(3) In a recent study \cite{Murphyetal2019} find that on average CCSN simulations obtain diagnostic energies that are lower than those observed. I predict that the inclusion of magnetic fields in CCSN simulations, with high resolution near the NS, will yield the observed energies, and up to $>10^{52} \erg$, as magnetohydrodynamics will lead to more collimated and more energetic jets, in particular with the inclusion of angular momentum in the pre-collapse core.  
Magnetohydrodynamical simulations of CCSNe are within reach, but require very large computational resources. Existing simulations (e.g. \citealt{Masadaetal2015, Mostaetal2015, ObergaulingerAloy2017, Obergaulingeretal2018}) do not include all necessary ingredients, such as pre-collapse perturbations and very high resolution near the NS. 

Overall, the conclusion of this study suggests that there is a smooth transition from significant role of neutrino heating in forming the jets to less significant at very energetic explosions. In particular, when the pre-collapse core is rapidly rotating the jets maintain a more or less fixed axis and might lead to super-energetic CCSNe \citep{GilkisSoker2016Super, Soker2017RAA}, where neutrino heating plays a smaller role.

\section*{Acknowledgments}
I thank Bernhard  M{\"u}ller for privately supplying some figures and for critical comments.  
I thank Amit Kashi for his crucial help in editing some figures and for helpful comments. I thank Avishai Gilkis and Jason Nordhaus for helpful comments.

\end{document}